\documentclass[times, twocolumn, trackchanges]{aastex63}
\usepackage{graphicx, graphics}
\usepackage{CJK}
\usepackage{amsmath, amssymb, bm}

\usepackage[nodayofweek]{datetime}
\newdateformat{monthyearday}{%
  \THEYEAR\ \monthname[\THEMONTH] \twodigit{\THEDAY}}
\usepackage{multirow}

\makeatletter
\patchcmd{\NAT@citex}
  {\@citea\NAT@hyper@{%
     \NAT@nmfmt{\NAT@nm}%
     \hyper@natlinkbreak{\NAT@aysep\NAT@spacechar}{\@citeb\@extra@b@citeb}%
     \NAT@date}}
  {\@citea\NAT@nmfmt{\NAT@nm}%
   \NAT@aysep\NAT@spacechar\NAT@hyper@{\NAT@date}}{}{}

\patchcmd{\NAT@citex}
  {\@citea\NAT@hyper@{%
     \NAT@nmfmt{\NAT@nm}%
     \hyper@natlinkbreak{\NAT@spacechar\NAT@@open\if*#1*\else#1\NAT@spacechar\fi}%
       {\@citeb\@extra@b@citeb}%
     \NAT@date}}
  {\@citea\NAT@nmfmt{\NAT@nm}%
   \NAT@spacechar\NAT@@open\if*#1*\else#1\NAT@spacechar\fi\NAT@hyper@{\NAT@date}}
  {}{}
\makeatother
\newcommand{\Qphi}{$\mathcal{Q}_\phi$}

\newcommand{\uat}[2]{\href{http://vocabs.ands.org.au/repository/api/lda/aas/the-unified-astronomy-thesaurus/current/resource.html?uri=http://astrothesaurus.org/uat/#1}{#2  (#1)}}
\newcommand{\affilCaltechAstro}{\affiliation{Department of Astronomy, California Institute of Technology, MC 249-17, 1200 East California Boulevard, Pasadena, CA 91125, USA; \url{ren@caltech.edu}}}

\received{2020 November 03}
\revised{2020 December 07}%\monthyearday\today}
\accepted{2020 December 08}
\acceptjournal{The Astrophysical Journal Letters}

\shorttitle{SAO 206462 Spiral Motion}
\shortauthors{Xie et al.}

\begin{document}
\pagenumbering{arabic}
\begin{CJK*}{UTF8}{gbsn}
\title{Spiral Arm Pattern Motion in the SAO~206462 Protoplanetary Disk}

\author[0000-0001-8184-5547]{Chengyan Xie  (谢承炎)}
\affiliation{Department of Astronomy, Xiamen University, 1 Zengcuoan West Road, Xiamen, Fujian 361005, China}

\author[0000-0003-1698-9696]{Bin Ren  (任彬)}
\affilCaltechAstro

\author[0000-0001-9290-7846]{Ruobing Dong  (董若冰)}
\affiliation{Department of Physics \& Astronomy, University of Victoria, Victoria, BC, V8P 1A1, Canada}

\author{Laurent Pueyo}
\affiliation{Space Telescope Science Institute (STScI), 3700 San Martin Drive, Baltimore, MD 21218, USA}

\author[0000-0003-2233-4821]{Jean-Baptiste Ruffio}
\affilCaltechAstro

%alphabet
\author[0000-0002-2853-3808]{Taotao Fang  (方陶陶)}
\affiliation{Department of Astronomy, Xiamen University, 1 Zengcuoan West Road, Xiamen, Fujian 361005, China}

\author[0000-0002-8895-4735]{Dimitri Mawet}
\affilCaltechAstro

\author[0000-0002-5823-3072]{Tomas Stolker}
\affiliation{Leiden Observatory, Leiden University, Niels Bohrweg 2, 2333 CA Leiden, The Netherlands}

\begin{abstract}
Spiral arms have been observed in more than a dozen protoplanetary disks, yet the origin of nearly all systems is under debate. Multi-epoch monitoring of spiral arm morphology offers a dynamical way in distinguishing two leading arm formation mechanisms: companion-driven, and gravitational instability induction, since these mechanisms predict distinct motion patterns. By analyzing multi-epoch $J$-band observations of the SAO~206462 system using the SPHERE instrument on the Very Large Telescope (VLT) in 2015 and 2016, we measure the pattern motion for its two prominent spiral arms in polarized light. On one hand, if both arms are comoving, they can be driven by a planet at $86_{-13}^{+18}$~au on a circular orbit, with gravitational instability motion ruled out. On the other hand, they can be driven by two planets at $120_{-30}^{+30}$~au and $49_{-5}^{+6}$~au, offering a tentative evidence ($3.0\sigma$) that the two spirals are moving independently. The independent arm motion is possibly supported by our analysis of a re-reduction of archival observations using the NICMOS instrument onboard the \textit{Hubble Space Telescope} (\textit{HST}) in 1998 and 2005, yet artifacts including shadows can manifest spurious arm motion in \textit{HST} observations. We expect future re-observations to better constrain the motion mechanism for the SAO~206462 spiral arms.
\end{abstract}


\section{Introduction}
More than a dozen young stars host spiral arms in their surrounding protoplanetary disks, however the formation mechanisms of the spirals are still under debate \citep{dong18b}. Two leading mechanisms predict distinct motion rates for these spirals: the spiral arms may be excited by companion(s), thus corotating with the companion(s) \citep[][]{kley12, dong15}, or produced by gravitational instability (GI), thus undergoing local Keplerian motion before gradually winding up and being destructed \citep{dong15gi, kratter16}. The true origin of spirals in nearly all systems is still in debate. On one hand, if they are produced by companions (e.g., stars, substars, planets), none of the predicted planetary drivers has been confirmed through direct imaging \citep[``missing planets'':][]{brittain20}. On the other hand, disk mass estimates under conventional assumptions suggest that most spiral systems are unlikely to be GI unstable \citep{dong18b}.
The discovery or non-detection of arm-driving companions serves as not only a test for spiral arm formation mechanisms \citep{dong16, wagner18, rosotti20, gonzalez20}, but also a proxy to the occurrence rate and formation mechanism of planets \citep{brittain20}. 

Measurement of pattern motion using multi-epoch observations offers a dynamical approach in distinguishing the two leading arm motion mechanisms and tracing the location of spiral-arm-driving planets \citep{ren20}. Using two epochs of observations separated by 5 years, \citet{ren20} perform pattern speed measurement for the two spiral arms in scattered light for the MWC~758 protoplanetary disk, and provide a dynamical evidence that they are simultaneously driven by one hidden planetary driver. Being one of the only two spiral-arm-hosting protoplanetary disks that have multi-epoch observations with VLT/SPHERE, here we analyze the motion of the spiral arms surrounding SAO~206462.

SAO~206462 (a.k.a., HD~135344B), a $12_{-6}^{+4}$~Myr old F4Ve star with a mass of $1.6_{-0.1}^{+0.1}~M_\sun$ \citep{garufi18}, is located in the Upper Cen star forming region at a distance of $135.0\pm0.4$~pc \citep{gaiaedr3}. Its submilimiter images show a large cavity within a radius of $45$~au \citep[e.g.,][]{andrews11, perez14, vandermarel16, cazzoletti18}, and its high-resolution scattered light image shows the existence of two prominent spiral arms \citep[e.g.,][]{muto12, garufi13, stolker16}. Using multi-epoch VLT/SPHRERE observations, \citet{stolker16} trace the spirals down to ${\sim}20$~au while observing shadowing effects that may originate from an inner disk, and \citet{stolker17} explain that the shadows can originate from localized perturbations in the inner disk. 

Despite theoretical and numerical attempts to explain the architecture of this system, the formation mechanisms for the two prominent arms surrounding SAO~206462 are unclear. On one hand, they can be driven by multiple planets \citep{muto12, stolker16} or a single planet \citep{bae16, dong17}. On the other hand, they can form through GI in massive disks, if the disk mass is ${\sim}25$--$50\%$ of the stellar mass \citep{dong15gi}. In this Letter, we analyze the multi-epoch VLT/SPHERE observations for the SAO~206462 system, and dynamically quantify the motion mechanism for its spirals arms for the first time. We describe our data reduction procedure in Section~\ref{sec-odr}, analyze the observations in Section~\ref{sec-ana}, discuss the findings in Section~\ref{sec-dis}, and summarize this Letter in Section~\ref{sec-sum}.

\section{Observation and Data Reduction}\label{sec-odr}

We obtain the SPHERE/IRDIS $J$-band dual-polarization imaging  observations of SAO~206462 in polarized light presented in \citet{stolker16, stolker17}. There is one observation in 2015 May, and four observations between 2016 May and 2016 June, establishing a 14~month temporal baseline for motion measurement. The 2015 observation uses apodizer APO1 (optimized for $4\lambda/D$ focal masks), and Lyot mask ALC1 (diameter: 145~mas; coronagraph combination name: {\tt N\_ALC\_YJ\_S}); the 2016 observations use apodizer APO1 and Lyot mask ALC2 (diameter: 185~mas; {\tt N\_ALC\_YJH\_S}). In all observations, the pixel scale is $12.25$~mas \citep{maire16}; the detector integration time is 32~s per frame. The total integration time spans from 17~min to 102~min, with the 2015 May 3 and 2016 May 4 observations being the two longest integrations (76.8~min and 102.4~min, respectively; see \citealp{stolker17} for the observation log).

\begin{figure*}[htb!]
	\includegraphics[width=\textwidth]{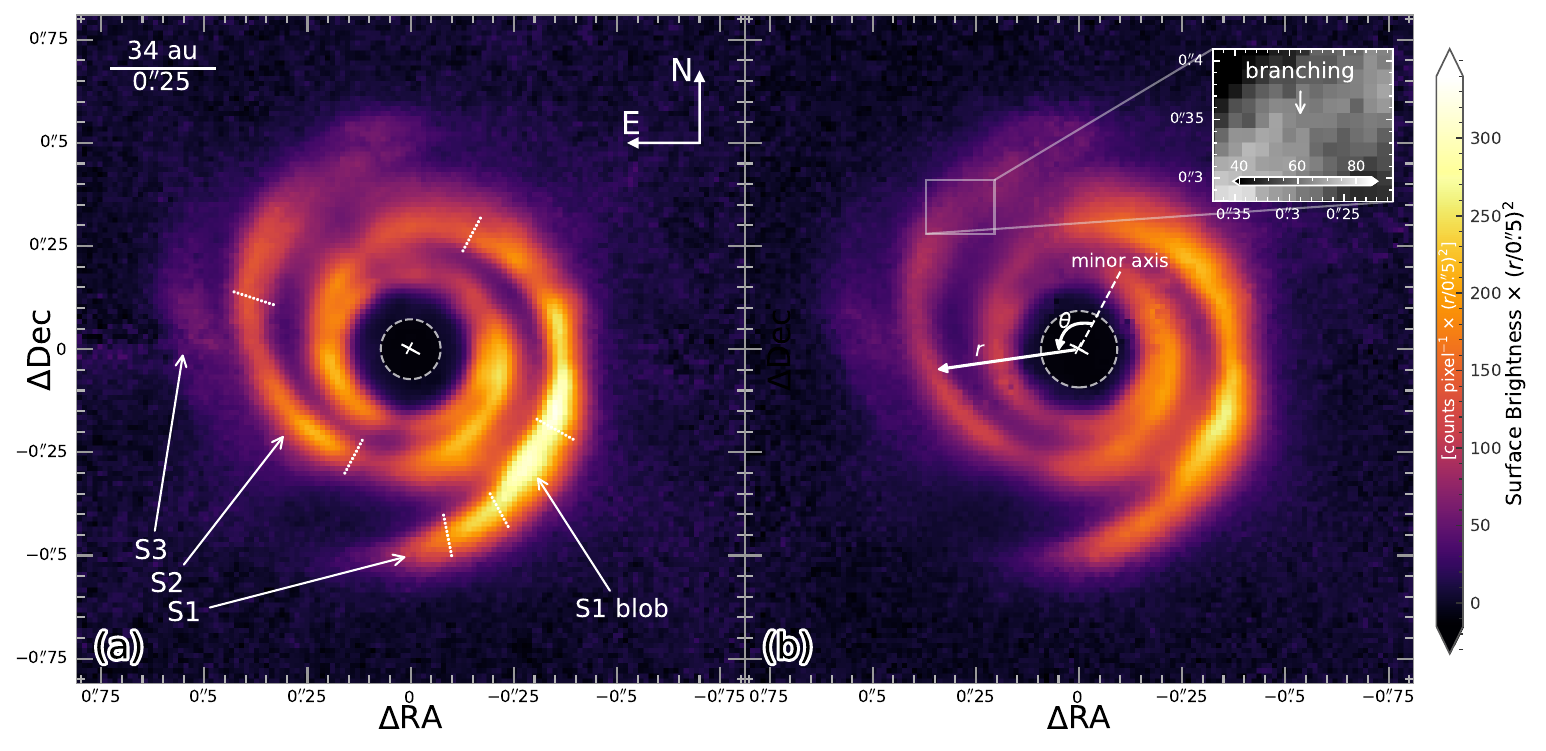}
    \caption{$r^2$-scaled \Qphi\ images of the SAO~206462 system on (\textbf{a}) 2015 May 3 and (\textbf{b}) 2016 May 4. The central dashed circles are the physical size of the coronagraphs, the cross is the location of the star, the longer side of the cross matches the major axis of the disk. Note: the color bars share the same units; $\theta$ and $r$ are measured in the disk plane; the dotted lines are the boundaries for motion measurement in Figure~\ref{fig2}.}
    \label{fig1}
    
    (The data used to create this figure are available.)
\end{figure*}

We reduce the five observations using {\tt IRDAP} \citep[][]{vanholstein20} with identical default parameters to minimize systematic offset, and analyze the output star-polarization-subtracted \Qphi\ files that trace the surface distribution of dust particles \citep{monnier19}. To minimize stellar illumination effects, we follow the \citet{ren20} procedure to scale the \Qphi\ images: we first compute the stellocentric distance, $r$, for each pixel and multiply its corresponding \Qphi\ value by $ ({r}/{r_0})^2$, where $r_0 = 0\farcs5$ (i.e., ``$r^2$-scaled''), assuming an inclination of $11^\circ$ from face-on, and a position angle of $62^\circ$ for the major axis of the disk \citep{dent05, perez14}. We present the $r^2$-scaled 2015 May 3 and 2016 May 4 images in Figure~\ref{fig1} and annotate the features following \citet{maire17}. For a complete set of all the images, we refer the readers to Figure~1 of \citet{stolker17}. We then deproject these $r^2$-scaled images to face-on views (i.e., the disk plane) and transform them to polar coordinates for spiral arm location measurement. In the interpolation procedure, we adopt the physically motivated cubic splines, which minimize the elastic energy for a system \citep{horn83}, implemented in the {\tt scipy.interpolate.interp2d} function in {\tt scipy} \citep{virtanen20}. 

\begin{figure*}[htb!]
    \centering
	\includegraphics[width=0.99\textwidth]{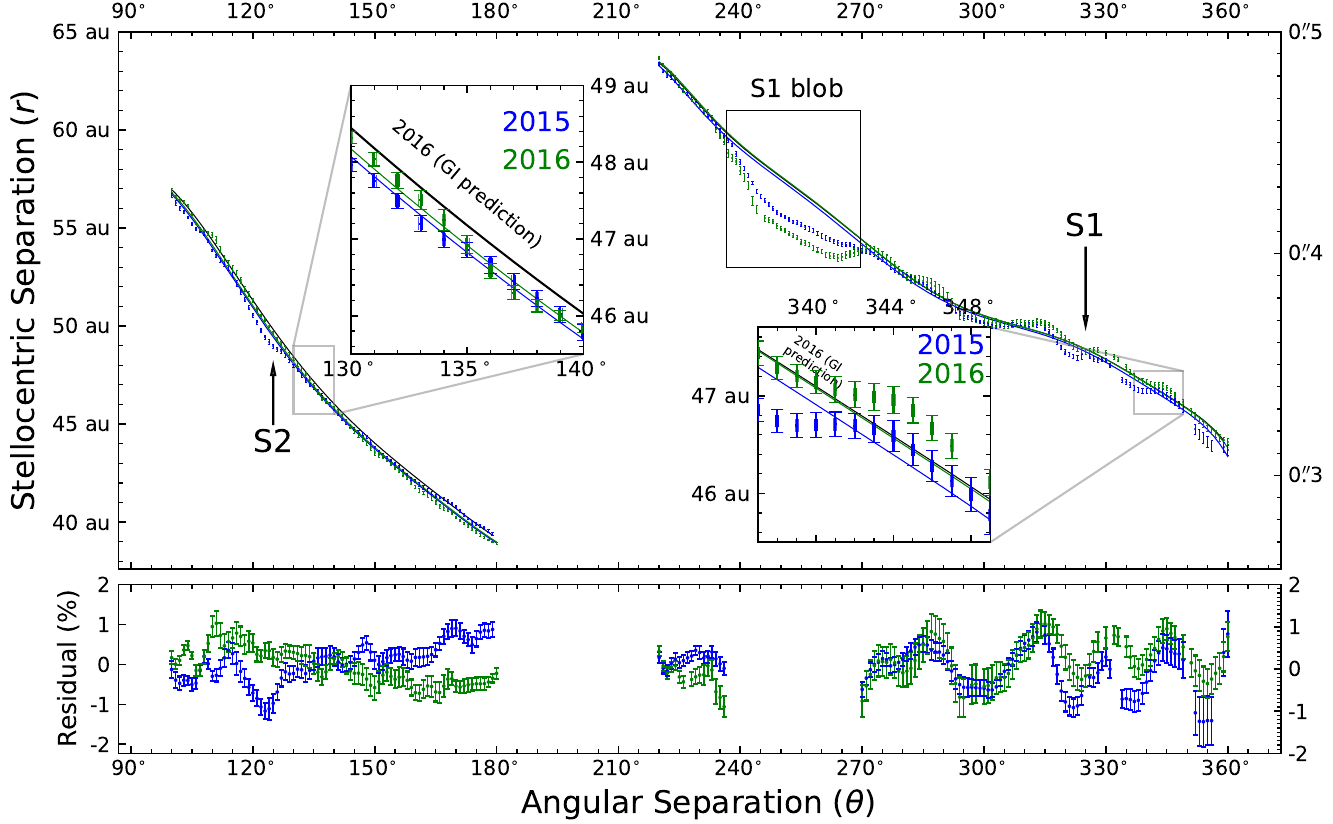}
    \caption{Morphology and independent motion fit for the S1 and S2 arms (error bars: arm location measurements; colored lines: fitted arm location driven by two individual planets; black lines: GI prediction of arm location around a $1.6~M_\sun$ central mass), the bottom panel shows the residuals of the selected points for motion measurement. Note: we do not include the S1 blob in the motion measurement. 
    }
    \label{fig2}
\end{figure*}

\section{Analysis}\label{sec-ana}
We use the two observations on 2015 May 3 and 2016 May 4 that provide a temporal separation of $1.00$~yr with the highest data quality for analysis. The exposure times of the rest of the observations are $17$~min or $34$~min \citep{stolker17}, which are a factor of more than twice shorter and thus provide compromised data quality. In addition, the three shorter observations constitute a temporal separation of less than 50 days, which provide shorter timeline for spiral motion. We discuss the contribution of these three short observations in Section~\ref{sec-dis-RE}.

\subsection{Spiral Location}
We measure the spiral locations after transforming the deprojected $r^2$-scaled \Qphi\ images into polar coordinates, where the horizontal axis is the counterclockwise angular deviation $\theta$ from the northwest semi-minor axis of the disk, and the vertical axis is the stellocentric distance $r$. 
For each $\theta$, we fit a Gaussian profile to its radial profile with {\tt scipy.optimize.curve\_fit} to obtain the peak location $r$ with an error\footnote{The errors in this Letter are $1\sigma$ unless otherwise specified.} of $\delta r$. 

When we inspect the radial profiles for the $\theta$ values, we notice multiple peaks or flat plateau in some regions in the S2 spiral, which is indicative of resolved and unresolved sub-spirals; we ignore these points to minimize their potential impact for location and subsequent speed measurement. See Figure~\ref{fig2} for the $(\theta, r\pm\delta r)$ measurements used for subsequent analysis.

\begin{deluxetable*}{l|l|cc|c}[thb!]
%\tablewidth{} 
% \setlength{\tabcolsep}{2pt}
\tablecaption{Pattern motion measurement for SAO~206462 spirals \label{tab1}}
\tablehead{ 
\multirow{2}{*}{Mechanism}  &  \multirow{2}{*}{Parameter} & \multicolumn{2}{c|}{Independent Fit} & Joint Fit\\\cline{3-5}
 					&						& S1 		& S2		&  S1 \& S2
 }
\startdata 
\multirow{3}{*}{Planet-Driven} & Rotation Rate (yr$^{-1}$) &$1\fdg32 \pm 0\fdg18$ & $0\fdg38\pm0\fdg13$ & $0\fdg57 \pm 0\fdg13$ \\
            & Driver Location\tablenotemark{a} (au) & $49_{-5}^{+6}$ & $120_{-30}^{+30}$ & $86_{-13}^{+18}$\\ 
            & Driver Orbital Period\tablenotemark{a} (yr) & $270_{-30}^{+50}$ & $950_{-240}^{+490}$ & $630_{-120}^{+190}$ \\ \hline
%\multirow{2}{*}{GI-Induction} & Rotation Rate\tablenotemark{b} (yr$^{-1}$) &$2\fdg46 \pm 0\fdg23$ &$0\fdg15\pm0\fdg14$ & $0\fdg46 \pm 0\fdg14$\\
%            & Inferred Central Mass\tablenotemark{c} ($M_\sun$) & $3.0\pm0.6$& $0.01_{-0.01}^{+0.03}$ & $0.10_{-0.05}^{+0.08}$
\multirow{2}{*}{GI-Induction} & Rotation Rate\tablenotemark{b} (yr$^{-1}$) &$\cdots$ &$\cdots$ & $0\fdg46 \pm 0\fdg14$\\
            & Enclosed Mass\tablenotemark{c} ($M_\sun$) & $\cdots$& $\cdots$ & $0.10_{-0.05}^{+0.08}$
            \enddata
\tablecomments{\tablenotetext{a}{The driver has a circular orbit along the midplane of the disk, its location is calculated for a  $1.6_{-0.1}^{+0.1}~M_\sun$ central star.}
\tablenotetext{b}{The rate for GI-induction is calculated for a location of $40$~au, multiply the rate by $\left(\frac{r}{40~{\rm au}}\right)^{-3/2}$ to obtain that for other locations.}}
\tablenotetext{c}{Enclosed mass within $39$~au, which is inferred from Keplerian motion.}
\end{deluxetable*}

\subsection{Spiral Motion}
To constrain the motion pattern for the spirals, we ignore
features that can bias our results. For example, the spirals joining the edge of the coronagraph (S1: $\theta \gtrsim 360^\circ$; S2: $\theta \gtrsim 180^\circ$), and the branching feature at the S2 tip (e.g., Figure~\ref{fig1} inset), may bias our spiral arm location measurement. The $r^2$-scaled surface brightness measurement of S3 is a factor of ${\sim}5$ lower than that of S2 in Figure~\ref{fig1}, and thus the influence from the merging of S2 and S3 should be less than ${\sim}20\%$. Therefore, we focus on $220^\circ \leq \theta \leq 360^\circ$ for S1, and $100^\circ \leq \theta \leq 180^\circ$ for S2. In our measurement, we additionally ignore the S1 blob (specifically, $237^{\circ}\leq\theta\leq269^{\circ}$) for S1, see Section~\ref{sec-s1-motion} for the justification. In Figure~\ref{fig2}, we present the selected data points with $1^\circ$ step for $\theta$; we also present the S1 blob for illustration purpose only. We denote these chosen angles in Figure~\ref{fig1} by projecting a three-dimensional setup of a disk to the sky plane \citep[Appendix~A of][]{ren19}.

%S1 220, 237, 269, 360; S2: 100, 180
% cos^2(\theta_{figure}) = 1/(cos^{-2}(\theta_{detector}-\theta_{PA}) + tan^2(\theta_{inc})) derived from there
%xOy angle from +x-axis on Figure 1 are then: 
% S1: 281.22, 298.79, 331.00, 422.00
% S2: 162.01, 242.00

\begin{figure*}[htb!]
	\includegraphics[width=\textwidth]{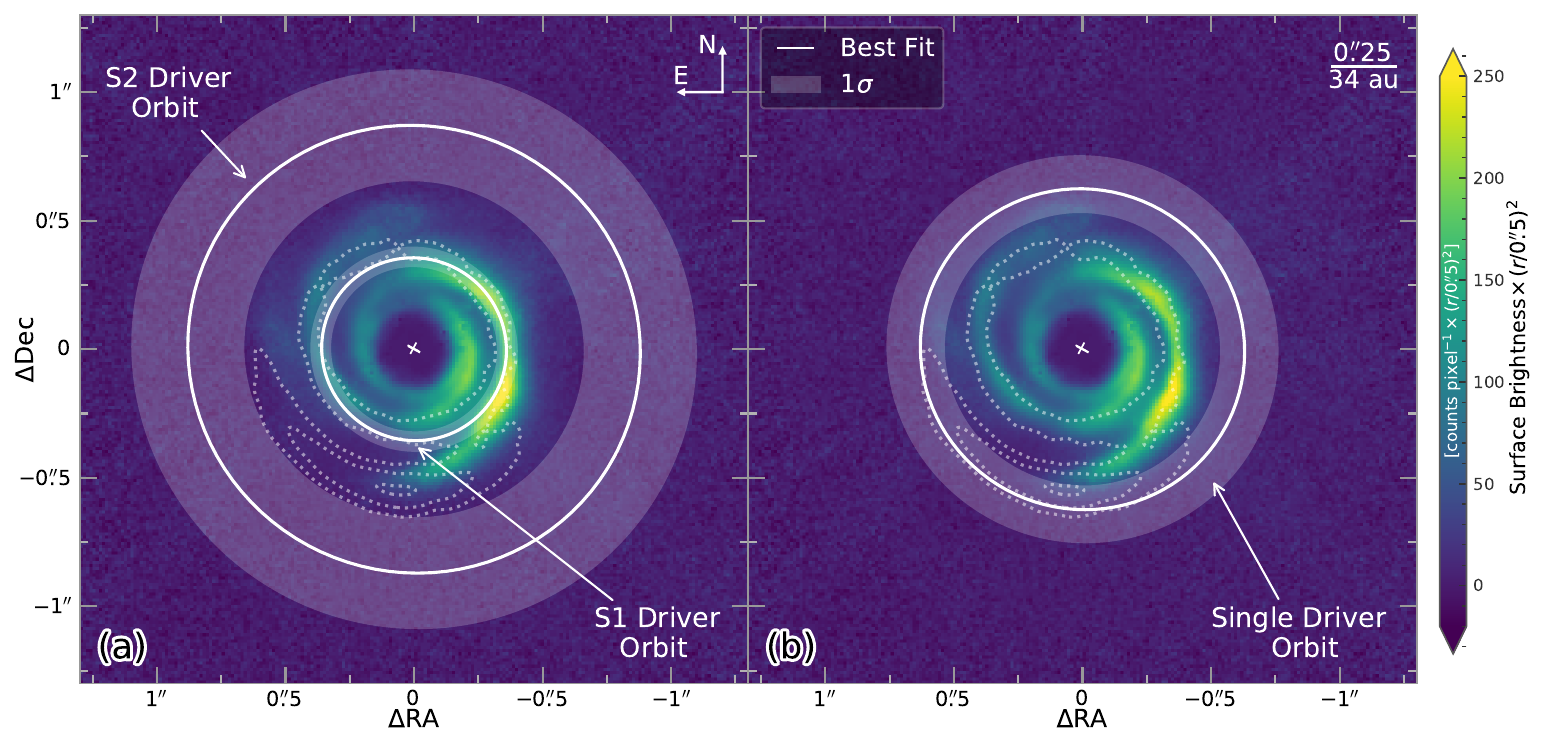}
    \caption{The best fit $\pm1\sigma$ orbits for assumed planetary drivers with circular orbits overlaid on ALMA contours \citep[$1.9$~mm; ][]{cazzoletti18} and $r^2$-scaled SPHERE \Qphi\ image ($1.2~\micron$). (\textbf{a}) double-driver scenario. (\textbf{b}) single-driver scenario.}
    \label{fig3}
\end{figure*}

We constrain the morphology and the angular movement between the two epochs for each arm under two hypotheses. A $ (\theta,r)$ pair will advance to a location of $(\theta+\Delta \theta,r)$ between the observations. In the GI-induction scenario, each part of the arm moves at the local Keplerian speed, $\Delta \theta \propto r^{-3/2}$.\footnote{Keplerian rotation is governed by the mass of the star and the disk enclosed at each radii $r$, i.e., $M_{r'<r}$, thus its speed falls off with radius slower than $r^{-3/2}$. As a first order approximation we do not take into account the radial dependence of $M_{r<r'}$ over the limited radial range concerned in our measurements, $43.9$--$64.1$~au for S1 and $39.1$--$57.3$~au for S2.}
In the companion-driven scenario, the entire arm corotates with its driver planet as a rigid body, %. In this scenario, 
thus $\Delta \theta = {\rm Constant}$ and traces the motion of the driver.

We fit $p$-degree polynomials to the $(\theta, r \pm \delta r)$
pairs in both epochs. Following \citet{ren20}, we use dummy variables as proxies to simultaneously obtain morphological parameters for a spiral and speed for pattern motion. Noticing that the two arms might be moving under different rates, we begin with independent fitting for them. In this Letter, positive pattern speed corresponds to counterclockwise rotation.

\subsubsection{Independent Motion}\label{sec-s1-motion}

Spiral arms are expected to be trailing features in protoplanetary disks. The orientation of the spirals in SAO 206462 indicates that the disk is rotating counterclockwise.
Nevertheless, we first obtain an S1 pattern speed of ${\sim}1^{\circ}$~yr$^{-1}$ clockwise under both the GI-induction and the planet-driven scenarios. Such a motion is to the opposite direction of the expected disk rotation inferred from spiral morphology. We notice that the result originates from the points at $237^{\circ}\leq\theta\leq269^{\circ}$, see Figure~\ref{fig2}. Such a region has been identified as a kink in \citet{stolker16} and the S1 blob in \citet{maire17}, which matches the location of a hypothesized forming planet based on arm morphology fitting \citep{muto12}. We thus exclude the S1 points at $237^{\circ}\leq\theta\leq269^{\circ}$, which have ${\gtrsim}1$~au residuals when we do not ignore them in our fitting, to minimize the impact from unresolved spirals around a forming planet  (e.g., the twist in \citealp{boccaletti20}) on our pattern motion measurement. 

The S1 pattern speed is $1\fdg32 \pm 0\fdg13 $~yr$^{-1}$ in the planet-driven scenario. Taking into account the $0\fdg08$ instrumental North uncertainty of SPHERE \citep{maire16}, we obtain a propagated uncertainty\footnote{The instrumental North uncertainty impacts the position angle measurement for all the data points towards the same direction, rather than randomly assigning uncertainties for different data points.} of $\sqrt{ [1.00~{\rm yr}\times(0\fdg13~{\rm yr}^{-1})]^2+2\times0\fdg08^2}/(1.00~{\rm yr})=0\fdg18$~yr$^{-1}$.  For a $1.6_{-0.1}^{+0.1}~M_\sun$ central star \citep{garufi18}, this corresponds to a driver located at $49_{-5}^{+6}$~au assuming a circular orbit.

The S2 pattern speed is $0\fdg38\pm0\fdg06 $~yr$^{-1}$ in the planet-driven scenario. Taking into account the instrumental uncertainty, the motion rate is $0\fdg 38\pm0\fdg13$~yr$^{-1}$. For a $1.6_{-0.1}^{+0.1}~M_\sun$ central star, this corresponds to a driver located at $120_{-30}^{+30}$~au assuming a circular orbit.

We summarize the planet-driven motion rates in Table~\ref{tab1}.  We do not further calculate the individual motion under GI given such a treatment is less physically motivated; instead, we perform a joint GI motion in Section~\ref{sec-joint-motion}.

\subsubsection{Comotion}\label{sec-joint-motion}
The symmetry of S1 and S2 suggests that they are also possibly comoving. On one hand, they can be simultaneously driven by GI that can trace the central mass \citep[e.g.,][]{dong15gi}. On the other hand, they can be simultaneously driven by a single planetary driver \citep[e.g.,][]{bae16, dong17}.

In the GI-induction scenario, the motion rate is $(0\fdg46 \pm0\fdg 08)\times\left(\frac{r}{40~{\rm au}}\right)^{-3/2}$~yr$^{-1}$. Taking into account the $0\fdg08$ instrumental uncertainty, we obtain $ (0\fdg46\pm 0\fdg14)\times\left(\frac{r}{40~{\rm au}}\right)^{-3/2}$~yr$^{-1}$. This rate corresponds to a combined mass of $0.10_{-0.05}^{+0.08}~M_\sun$ for the central star and the inner disk, which is one order of magnitude smaller than the current mass estimate of the star \citep[$1.6_{-0.1}^{+0.1}~M_\sun$;][]{garufi18}. We thus do not favor the GI-induction scenario for the comotion of the two spirals.

In the planet-driven scenario, the pattern speed is $0\fdg57\pm0\fdg07 $~yr$^{-1}$. Taking into account the instrumental uncertainty, the rate is $0\fdg 57\pm0\fdg13$~yr$^{-1}$. For a $1.6_{-0.1}^{+0.1}~M_\sun$ star, it corresponds to a driver at $86_{-13}^{+18}$~au assuming a circular orbit.

\section{Discussion}\label{sec-dis}
\subsection{Morphological Fitting}
We obtain the best-fit $p$-degree polynomial description, $r(\theta) = \sum_{j=0}^p c_j \theta^j$ where $p\in\mathbb{N}$ and $c_j\in\mathbb{R}$ is the coefficient for the $j$th term, of the spirals by minimizing the Schwarz information criterion \citep{schwarz78} that penalizes excessive use of parameters. For the S1 arm, the best-fit $p=8$; the S2 arm, $p=5$. These best-fit $p$ parameters apply to both the GI-induction and the planet-driven scenarios.

\subsection{Robustness Estimation}\label{sec-dis-RE}
We compare our motion measurement with a cross-correlation analysis of disk images \citep[e.g.,][]{ren18}. For the selected regions in Figure~\ref{fig1}, the best fit motion rate based on cross-correlation is $-1\fdg2\pm57\fdg9$~yr$^{-1}$ for S1, $1\fdg1\pm43\fdg8$~yr$^{-1}$ for S2, and $-2\fdg2\pm56\fdg7$ for both. These rates, which report the motion in the planet-driven scenario, are dominated by shadowing effects since the most of the disk in 2016 is ${\sim}0.7\times$ the brightness in 2015 (with an exception of the northwest S1 arm: ${\sim}1.3\times$) in Figure~\ref{fig1}. In addition, the uncertainty from cross-correlation analysis traces the broadening of the signals \citep{tonry79},  thus the width of the spirals along the radial direction here, which is less informative on the real motion of the spirals. We therefore do not adopt the results from cross-correlation analysis. In this Letter, instead, we approximate the dust distribution for each angle with a Gaussian profile to locate the spines for the spiral arms, since we do not expect shadows to affect the radial distribution of dust particles. We note that an eccentric driver in the  \citet{calcino20} simulation may drive the spiral arm motion differently, however the corresponding arm motion has not been characterized.

We have assumed that the disk is infinitely thin in our deprojection procedure. Nevertheless, \citet{andrews11} report for SAO~206462 an aspect ratio, which is defined as the ratio between vertical scale height and radial separation (i.e., $h/r$), of $0.096 \left(\frac{r}{100~{\rm au}}\right)^{0.15}$ using the Submilimiter Array at 880~\micron. We use {\tt diskmap} \citep{stolker16b} for the deprojection of the system to address such effects, and find that in the planet-driven scenario, the S1 motion is $1\fdg12 \pm 0\fdg12$ yr$^{-1}$, the S2 motion $0\fdg38 \pm 0\fdg07$ yr$^{-1}$, and the comotion $0\fdg57 \pm 0\fdg07$ yr$^{-1}$; all of these are consistent with our previous results within $1\sigma$. In addition, although millimeter observation traces a different layer of dust from the scattered light observations, we expect that the impact from such a difference  is not important given the low inclination of this system (e.g., see \citealp{ren20} for their experiment on the impact of disk flaring).

We have used all data points in Figure~\ref{fig2} except the S1 blob in our motion analysis. To address possible bias from individual data pairs, we randomly discard 20\% of the pairs and repeat the motion analysis procedure for $10^4$ times. We find that for S1, the best fit rotation rate for the driver is $1\fdg32\pm0\fdg07$~yr$^{-1}$, $0\fdg38\pm 0\fdg06$~yr$^{-1}$ for S2, and $0\fdg57\pm 0\fdg04$~yr$^{-1}$ for comotion; all within $1\sigma$ from our initial measurements.

We have only used the two observations that have the longest exposure times for motion analysis. To address the contribution from the three shorter observations, we repeat the measurement using all five epochs. We exclude the third epoch due to its compromised data quality with a $17$~min exposure and a seeing larger than $2''$. For S1, the four-epoch result under planet-driven scenario is $1\fdg33 \pm 0\fdg12$ yr$^{-1}$, which is consistent with the previous two-epoch result within $1\sigma$. For S2, we obtain large uncertainties in the Gaussian fit for spiral arm location measurement. What is more, we could not properly approximate the data points at $150^\circ\lesssim\theta\lesssim180^\circ$ using Gaussian profiles in the two epochs with $34$~min exposures. We notice that the S2 arm is ${\sim}2$ times fainter than S1 in Figure~\ref{fig1}, we thus conclude that a $34$~min exposures is not sufficient in capturing the S2 arm with high data quality. In addition, we note that these shorter exposures can only establish a $40$~day timeline, which is a factor of $9$ less than the longer exposures in Section~\ref{sec-ana}. Therefore, we do not include the three short exposures in our analysis.

{We investigate the robustness of our measurements using the two $34$~min observations on 2016 June 22 and 30 that establish a $7.9$ day separation. After propagating the instrumental North uncertainty, we obtain that the angular motion rates under all planet-driven scenarios are $0^\circ\pm6^\circ$~yr$^{-1}$, or $0\fdg00\pm0\fdg13$ between the observations. The planet-driven arm motion in Table~\ref{tab1} during this $7.9$-day period is expected to range from $0\fdg01$ to $0\fdg03$, which is within the $1\sigma$ interval of the estimate using the 2016 June data. In addition to instrumental North and statistical uncertainties, the uncertainty in our measurements may originate from effects including (but not limited to) compromised data quality with $34$~min observations, random noise, and shadows from inner disks. Specifically, shadows with 1 week period can trace down inner disks at $0.1$~au. Nonetheless, we cannot properly decompose these effects until time series monitoring of the system is available.}

\subsection{Independent Motion}\label{sec-dis-arm-motion}
For the S1 arm, its motion is consistent with being driven by a planet when we exclude the S1 blob region. The driver is located at $49_{-5}^{+6}$~au assuming a circular orbit, which coincides with the stellocentric radius of the S1 blob. In comparison, on one hand, \citet{muto12} fit the spiral morphology and report a theorized driver at $53$~au, while \citet{stolker16} obtain $24$~au assuming the driver is located within the cavity：our morphological fit using identical observation and method, but without such a requirement returns loose constraints for the S1 driver -- the position angle is $40^\circ\pm100^\circ$, the stellocentric separation is $19\pm15$~au. On the other hand, \citet{maire17} identify the \citet{muto12} driver region as a bright S1 blob, and report a redder spectrum for the S1 blob than the spirals. Despite these results, we cannot constrain the position angle for the dynamically measured driver in this Letter, and thus we do not try to over-interpret the findings here. Nevertheless, such a location overlaps with the ALMA millimeter ring \citep[see Figure~\ref{fig3}, e.g.,][]{vandermarel16, cazzoletti18}. This overlapping could be explained by planet-disk interaction when two conditions -- low mass planet and low disk viscosity -- are met \citep[e.g.,][]{facchini20}.

For the S2 arm, its motion is consistent with being driven by a planet at $120_{-30}^{+30}$~au assuming a circular orbit. This agrees with predictions based on morphological estimates \citep[e.g.,][]{muto12, dong17, bae16}. Our morphological analysis based on static images for S2 returns a position angle of $20^\circ\pm2^\circ$ and location of $75\pm2$~au for the driver; yet we caution that the ignored regions, as well as the possibility of two interacting planets and thus spirals, may change the results.

Assuming S1 and S2 are driven by individual planets with circular orbits, we present the semi-major axes for the two hypothesized arm-driving planets in Figure~\ref{fig3}. Comparing the motion rates for the two arms under this scenario, we obtain a $3.0\sigma$ difference, which offers a tentative evidence that the two spirals are moving independently. Nonetheless, the colocation of the ALMA ring and the S1 driver from motion measurement requires a less massive planet  \citep[e.g.,][]{facchini20}, which is in tension with the expectation that spiral arms with high arm-to-disk contrast should be excited by massive planets \citep[e.g.,][]{dong17}. To further evaluate the difference between the spiral motion rates, we expect that a re-observation of the system after year 2020, which will establish a ${>}5$~yr timeline for motion measurement, is necessary \citep[e.g.,][]{ren20}.

\subsection{Driver Constraints}

We obtain the direct imaging constraints on the mass of the S2 driver with hot-start evolutionary models (i.e., Sonora, Bobcat; M.~Marley et al., in preparation) using $2400$~s of Keck/NIRC2 $L'$-band archival observation on 2016 May 27 (Program ID: C264N2, PI: D.~Mawet). We pre-process the data following \citet{Xuan18}, and subtract the stellar point spread function using a principal-component-analysis-based speckle subtraction method with a matched filter \citep[FMMF;][]{ruffio17}. We adopt the $W1$ magnitude of SAO~206462 \citep[$5.41\pm0.05$; CatWISE2020: ][]{eisenhardt20} as its $L'$-band magnitude, then follow \citet{ruffio18} to transform the Gaussian-distributed S2 pattern speed to planetary mass limit (with speed boundaries being the $2\sigma$ lower limit of S2 motion and the best fit of S1 motion). We obtain a planet-to-star flux ratio upper limit of $1.5\times 10^{-4}$ in $L'$-band (i.e., $\Delta L' = 9.6$) at $97\%$ confidence level. Adopting an age of $12_{-6}^{+4}$~Myr \citep{garufi18}, we calculate a mass upper limit of $13~M_{\rm Jupiter}$. This upper limit is consistent with the driver mass in theoretical predictions \citep[$5$--$15~M_{\rm Jupiter}$: ][]{bae16, dong17}.

A driver on a highly eccentric orbit may excite spiral arms \citep[e.g., eccentricity $e=0.4$ for the MWC~758 system: ][]{calcino20}. Nevertheless, the corresponding pattern motion characteristics have not been characterized yet. We therefore only discuss the impact of less eccentric drivers that do not trigger wiggles or bifurcations \citep[$e\lesssim0.2$: ][]{li19, muley19}. When $e=0.2$, the S2 driver has a possible range of $102$~au to $118$~au, which is consistent with the estimated $1\sigma$ uncertainty in Table~\ref{tab1}.

\subsection{Independent Motion or Comotion?}
The motion rates of the two spirals differ by $3.0\sigma$, yet the symmetry of the two spirals calls for a single planet driver \citep[e.g.,][]{bae16, dong17}. In comparison with the double-planet scenario where we obtain a $\chi^2$ value of $1617$ assuming independent Gaussian noise, we obtain $\chi^2=1662$ in the single-planet scenario. The $\chi^2$ difference between the two scenarios is $\Delta\chi^2 = 1662-1617=45$. We adopt the Akaike information criterion (AIC) and Schwarz information criterion (SIC), both of which penalize excessive use of free parameters, to compare the two scenarios. We obtain difference of $\Delta{\rm AIC} = 43$ and $\Delta{\rm SIC} = 39$, both are larger than the classical threshold of $10$ \citep[e.g.,][]{kass95}. However, correlated noise could decrease the difference in the $\chi^2$ (consequently the $\Delta{\rm AIC}$ and $\Delta{\rm SIC}$ values), we thus do not distinguish the two scenarios here. 

In addition to statistical noise, alternative spiral formation mechanisms could bias our motion measurement. On one hand, shadows, which trace the motion of the inner disk that is under the influence of local dust dynamics in SAO~206462 \citep{stolker17}, can affect the formation of spirals \citep{montesinos18} and thus impact the motion of spirals. On the other hand, an eccentric driver \citep[e.g.,][]{calcino20} could change the spiral motion pattern and fitting results. We thus do not attempt to conclude on the number of planetary drivers, and instead present the possible orbits for both planet-driven scenarios in Figure~\ref{fig3}.

\begin{figure}[htb!]
	\includegraphics[width=0.48\textwidth]{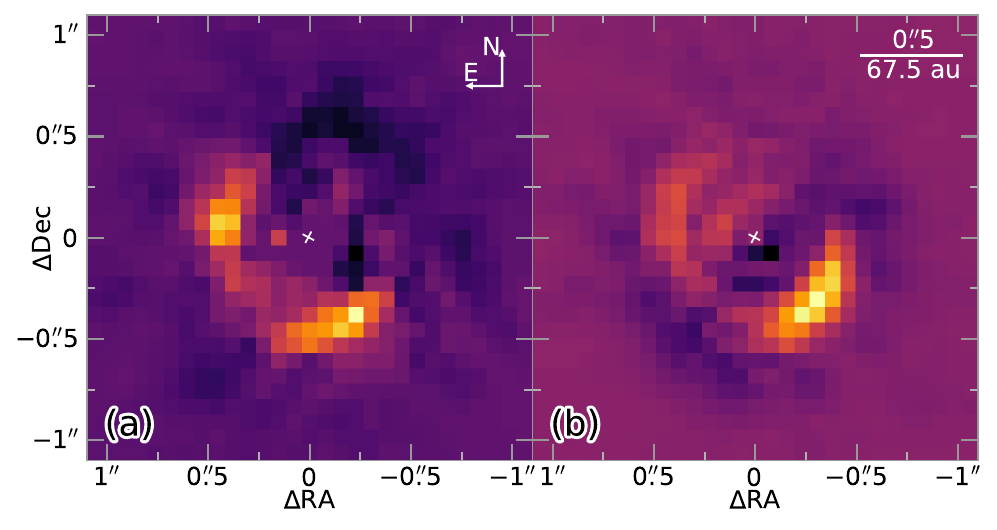}
    \caption{Reduced \textit{HST}/NICMOS observations of SAO~206462 in arbitrary units, presented in linear scale. (\textbf{a}) 1998 August 22, the F160W filter (${\sim}1.6~\mu$m). (\textbf{b}) 2005 March 24, F110W (${\sim}1.1~\mu$m). The motion of both spirals is possibly consistent with the independent motion in Table~\ref{tab1}. However, in addition to instrumental and reduction artifacts, the apparent motion of S1 may rise from the counterclockwise rotation of a shadow between the observations (i.e., from ${\sim}3$ o'clock to ${\sim}1$ o'clock).
    }\label{fig4}

    (The data used to create this figure are available.)
\end{figure}

We query and reduce available observations of the SAO~206462 system using \textit{HST}/NICMOS in 1998 (F160W filter, PropID: 7857, PI: A.-M.~Lagrange) and 2005 (F110W filter, PropID: 10177, PI: G.~Schneider) in \citet{grady09}. The apparent motion in Figure~\ref{fig4} is possibly consistent with independent motion. Nevertheless, we note that the NICMOS observations can be dominated by speckle noise, the NICMOS pixel size of $75.65$~mas is ${\sim}6$ times that of the SPHERE pixel, and that our principal-component-analysis-based data reduction method \citep{soummer12} can alter the morphology of the spirals in reference differential imaging. What is more, and most importantly, moving shadows may manifest spurious motion between the two observations -- specifically, the shadow at ${\sim}3$ o'clock in 1998 may have moved to ${\sim}1$ o'clock in 2005, thus causing spurious motion \citep[e.g.,][]{debes17}. Therefore, we do not perform motion measurement by combining these observations as in \citet{ren18}, nor do we distinguish between the independent motion and comotion mechanisms in this Letter.

\section{Summary}\label{sec-sum}
We have analyzed the spiral arm motion for the SAO~206462 protoplanetary disk system using five SPHERE observations in $J$-band polarized light. By comparing the two observations that have the longest exposures and constitute a temporal separation of $1.00$~yr, we measure the motion rates of the two major spiral arms, S1 and S2.

When we fit the motion for the spirals individually, S1 and S2 can be driven by planets with circular orbits at $49_{-5}^{+6}$~au and $120_{-30}^{+30}$~au, respectively. This offers a $3\sigma$ tentative evidence that spiral arms can move independently in one system. The orbits of these planetary drivers are consistent with some morphological fitting of spirals (e.g., S1 and S2: \citealp{muto12}, S2: \citealp{stolker16}). Although this is possibly consistent with our re-reduction of archival observations using \textit{HST}, we emphasize that artifacts including data reduction, instrumental instability, and shadows could result into spurious S1 motion with \textit{HST}. We therefore recommend follow-up SPHERE observations to better constrain the individual arm motion rates.

When we fit the motion for the spirals simultaneously, S1 and S2 can be driven a planetary driver on a circular orbit at $86_{-13}^{+18}$~au, or they are undergoing GI motion surrounding a central mass of $0.10_{-0.05}^{+0.08}$~$M_\sun$. The inferred central mass, which is a combination of the central star and the inner disk, under GI-induction is not consistent with the central star mass estimate of the star \citep[$1.6_{-0.1}^{+0.1}$~$M_\sun$;][]{garufi18}. The single-planet-driven result is consistent with the theoretical single-driver studies \citep[$100$--$120$~au; e.g.,][]{bae16, dong17} within $2\sigma$. 

We do not distinguish between the double-planet and the single-planet scenarios here given the existence of correlated noise, shadows that can impact spiral formation \citep[e.g.,][]{montesinos18}, and possible eccentric driver(s) in this system. Nevertheless, with our initial orbital constraints, such spiral-arm-driving planets are ideal targets for direct imaging using Keck/NIRC2, VLT/ERIS, and the \textit{James Webb Space Telescope}. We expect that a re-observation of the SAO~206462 system after 2020 using VLT/SPHERE will establish a ${>}5$~yr temporal baseline for motion measurement, which thus can not only help distinguish the two planet-driven scenarios, but also better constrain the semi-major axis for the planetary driver(s). 

\facilities{VLT:Melipal (SPHERE), Keck:II (NIRC2), \textit{HST} (NICMOS)}

\software{{\tt IRDAP} \citep{vanholstein20}, {\tt diskmap} \citep{stolker16b}, {\tt scipy} \citep{virtanen20}}

\acknowledgements 
We thank the anonymous referee for their suggestions that increased the clarity and robustness of this Letter, and Jaehan Bae for useful discussions. T.F.~and C.X.~are supported by the National Key R\&D Program of China No.~2017YFA0402600, project S202010384487 XMU Training Program of Innovation and Enterpreneurship for Undergraduate, and NSFC grants No.~11525312, 11890692. R.D.~acknowledges financial support provided by the Natural Sciences and Engineering Research Council of Canada through a Discovery Grant, as well as the Alfred P.~Sloan Foundation through a Sloan Research Fellowship. This research is partially supported by NASA ROSES XRP, award 80NSSC19K0294. Based on observations collected at the European Organisation for Astronomical Research in the Southern Hemisphere under ESO programmes 095.C-0273 (A), 097.C-0702 (A), 097.C-0885 (A), and 297.C-5023 (A). Based on observations made with the NASA/ESA \textit{Hubble Space Telescope}, obtained from the data archive at the Space Telescope Science Institute. STScI is operated by the Association of Universities for Research in Astronomy, Inc.~under NASA contract NAS 5-26555. Some of the data presented herein were obtained at the W.~M.~Keck Observatory, which is operated as a scientific partnership among the California Institute of Technology, the University of California and the National Aeronautics and Space Administration. The Observatory was made possible by the generous financial support of the W.~M.~Keck Foundation. The authors wish to recognize and acknowledge the very significant cultural role and reverence that the summit of Maunakea has always had within the indigenous Hawaiian community.  We are most fortunate to have the opportunity to conduct observations from this mountain.

\bibliography{refs}
\end{CJK*}
\end{document}